\documentclass[journal]{IEEEtran}
\usepackage{amssymb}
\usepackage{amsmath}

\usepackage{array}

\usepackage[dvipsnames]{xcolor}	

\usepackage{pgfplots}
\usepackage{Box}
\usepackage{RightBandedBox}
\usetikzlibrary{spy}
\usepackage{algorithm, algpseudocode}
\usepackage{enumerate}
\usepackage{graphicx}
\usepackage{bm}

\usepackage[nospace]{cite}

\usetikzlibrary{quotes,arrows.meta}
\usetikzlibrary{positioning}

\def\edgecolor{rgb:blue,4;red,1;green,4;black,3}
\newcommand{\midarrow}{\tikz \draw[-Stealth,line width =0.8mm,draw=\edgecolor] (-0.3,0) -- ++(0.3,0);}

\usetikzlibrary{3d} 

\def\ConvColor{rgb:yellow,5;red,2.5;white,5}
\def\BatchNormColor{rgb:green,5;black,2.5;white,5}
\def\PoolColor{rgb:orange,2.5;black,5;white,2.5}
\def\DropoutColor{rgb:blue,2.5;black,5;white,2.5}

\def\SigmoidColor{rgb:red,1;black,0.3}

\def\FcColor{rgb:purple,5;black,2.5;white,5}

\onecolumn

\begin{document}

\title{Fast approximate reciprocal approximations for iterative algorithms\\
	{\thanks{{The financial support by the Austrian Federal Ministry for Digital and Economic Affairs, the National Foundation for Research, Technology and Development and the Christian Doppler Research Association is gratefully acknowledged.\newline
This work has been supported by the COMET-K2 Center of the Linz Center of Mechatronics (LCM) funded by the Austrian federal government and the federal state of Upper Austria.	\newline			
				 M. Lunglmayr is with the Institute of Signal Processing, Johannes Kepler University Linz, 4040 Linz, Austria (e-mail: michael.lunglmayr@jku.at).\newline O. Ploder is with the Christian Doppler Laboratory for Digitally Assisted RF Transceivers for Future Mobile Communications, Institute of Signal Processing, Johannes Kepler University Linz, 4040 Linz, Austria (e-mail: oliver.ploder@jku.at).}}}}
\author{Michael~Lunglmayr,~\IEEEmembership{Member,~IEEE,} Oliver Ploder,~{Student Member,~IEEE}}

\maketitle

\begin{abstract}
The reciprocal function, $1/x$, is important for many real-time algorithms. It is used in a large variety of algorithms from areas ranging from iterative estimation to machine learning. Many of these algorithms are iterative in nature and require the online computation of the reciprocal. Such an iterative structure often prevents effective use of pipelining for implementation of the reciprocal. For this reason, a reciprocal algorithm requiring only a low amount of clock cycles is desired. Many real-time algorithms, often being of approximate nature, can tolerate the use of only an approximate solution of the reciprocal. 
 For this reason, we present a low complexity non-iterative approximation of the reciprocal function. This approximation can be calculated using only combinatorial logic. We present synthesis results showing that the proposed approach can be implemented with low area requirements at high clock frequencies. We analytically describe the error of the approximation and show that by optimizing a constant value used in the approximation, different variants with different error behaviors can be obtained. We furthermore present performance results of application examples that, when using our proposed method, show only negligible performance degradation compared to when using the exact reciprocal function, demonstrating the versatility of our proposed approach. 
\end{abstract}


\IEEEpeerreviewmaketitle

\section{Introduction}
The reciprocal is an important mathematical function for real-time implementation of algorithms in areas such as adaptive signal processing or machine learning. Example applications are in the online calculation of a step-size or learning rate in iterative or adaptive algorithms\cite{haykin, OLBILMS, ISCAS2017} or the calculation of activation functions of neural networks \cite{GradDescTraining}.
These applications have in common that the reciprocal function is used inside iterations of an algorithm, and is potentially used multiple times in a hardware architecture in a parallelized structure. For this reason, an architecture would be beneficial that can be performed non-iteratively with only a small number of clock cycles (potentially within only one clock cycle) and only requires a low amount of hardware. The benefit of implementing the reciprocal with only a small number of clock cycles stems from the fact that, in low complexity iterative algorithms, a whole iteration can often be implemented with a small number of clock cycles. Because in iterative algorithms the operand of the reciprocal function is often not known beforehand, the possibilities for improvement by pipelining is often limited. The number of clock cycles required for calculating the reciprocal then adds to the number of clock cycles required for an iteration. 
Although a large variety of algorithms for the implementation of the reciprocal function have been proposed in the literature before, the proposed approaches typically favor high precision over a low number of clock cycles.
Besides classical algorithms for implementing a division, often based on digit recurrence algorithms (e.g. see \cite{DivisionOverView} and the references therein), algorithms proposed for real-time implementation often rely on iterative algorithms based on the Newton method \cite{raps1,raps2,ScalingLessNewton} or Goldschmidt's algorithm \cite{Gold1, Gold2, Goldschmidt}. Other approaches use lookup tables \cite{VLSICell} or combine lookup tables with iterative methods \cite{tab1, HighSpeedTC}. For these proposed methods, the first design goal typically is to obtain a high precision reciprocal function. Latency and complexity is optimized subsequently while maintaining a precision as high as possible. 

In this work, we follow a different approach. We observed that for many iterative algorithms an approximate algorithm for the reciprocal function performs sufficiently well (given, that the error is within certain bounds as we will describe below). But often it is of crucial importance to perform the calculation of the (approximate) reciprocal with a small number of clock cycles. We therefore propose a method based on a piecewise linearization of the $1/x$ function using powers of two as linearization points. The method is non-iterative and does not use a lookup table. It allows performing an approximation of the reciprocal function with a leading one detector, two shift operations, and one subtraction. We show how to analytically optimize a parameter of this approximation such that the approximation is either always below the exact curve $1/x$, above the exact curve, or it has an average relative error, compared to the exact curve, of zero. We furthermore analytically describe the error and derive bounds on maximum relative error for different variants of the approximation.
We furthermore present a monotonicity preserving extension that allows obtaining a maximum relative error of less than $6\%$ by using one additional comparison operation.

We present hardware architectures demonstrating that the proposed method can be implemented in one clock cycle with low area requirements and high clock frequencies. We furthermore present synthesis results demonstrating the performance when introducing intermediate registers between the combinatorial blocks (as it is often done to further increase the clock frequency).

\section{Piecewise linear interpolation of $1/x$ }
The idea of representing $1/x$ as a piecewise linear function has been already proposed in \cite{PieceWise}.  There the breakpoints (connecting points of the line segments) of the linear segments have been optimized to minimize the approximation error. This approximation was then used in a bitstream-based architecture to approximate the $1/x$ function. 

In this work, we will use a different approach. Although we also approximate the reciprocal by a piecewise linear function, the breakpoints of the segments have not been chosen with the aim of minimizing the approximation error but with the aim of designing a method with low complexity. For simplicity, in this work we will assume that we only calculate $1/x$ for values $x>1$. 

\begin{figure}[h]
	\centering
	\includegraphics[width=0.75\columnwidth]{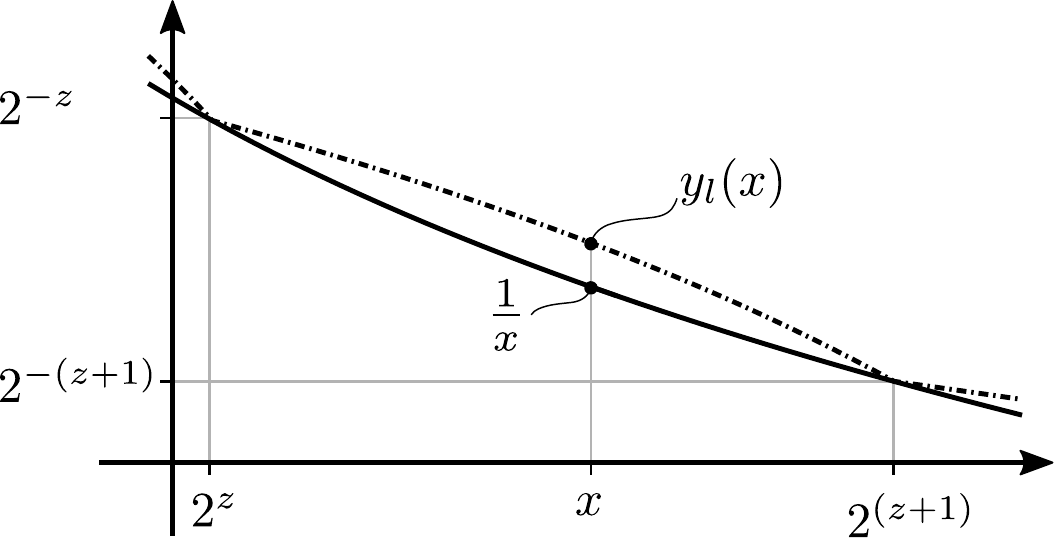}
	\caption{Piecewise linearization of $1/x$.}
	\label{fig:piecewise_lin}
\end{figure}

The main idea is as follows. We only use powers of two as breakpoints for all line segments. For every line segment we use $(2^{z},2^{-z})$ as start point and $(2^{(z+1)},2^{-(z+1)})$ as endpoint, with $z \in \{0,1,2, \ldots \}$. 
Fig.~\ref{fig:piecewise_lin} graphically shows this approximation. 
The idea is to use the next value lower or equal to $x$ as $2^z = 2^{\lfloor \text{log}2(x) \rfloor}$ and the next value greater or equal to $x$ as $2^{z+1}= 2^{\lfloor \text{log}2(x) \rfloor+1}$. With these linearization endpoints, for each value $2^z \leq x  < 2^{z+1}$ the piecewise linear representation can be described by 
\begin{align}
(1-a) 2^{-z} + a 2^{-(z+1)} = (2-a) 2^{-(z+1)},
\label{eqn:seqment}
\end{align}
with the linear combination factor $0 \leq a \leq 1$. For a given value $x$, the corresponding value $a$ can be calculated as:
\begin{align}
a = \frac{x-2^{-z} }{2^{-(z+1)}- 2^{-z}} = x 2^{-z} - 1
\end{align}
Using $a$ in (\ref{eqn:seqment}) results in the approximate function $y_l$:
\begin{align}
y_l(x,C) = (C-x 2^{-z}) 2^{-(z+1)},
\label{eqn:interpol}
\end{align}
where the constant $C=3$ results from the above derivation. In this paper, we always use\footnote{because the variable $z$ depends on $x$ we ommit the depency on $z$ in $y_l(x,C)$} $z=\lfloor\text{log}_2 x\rfloor$.
Using a more general constant $C$ allows to change the characteristics of the approximation as we will describe below.

\begin{figure}[h]
	\centering
	\includegraphics[width=0.35\columnwidth]{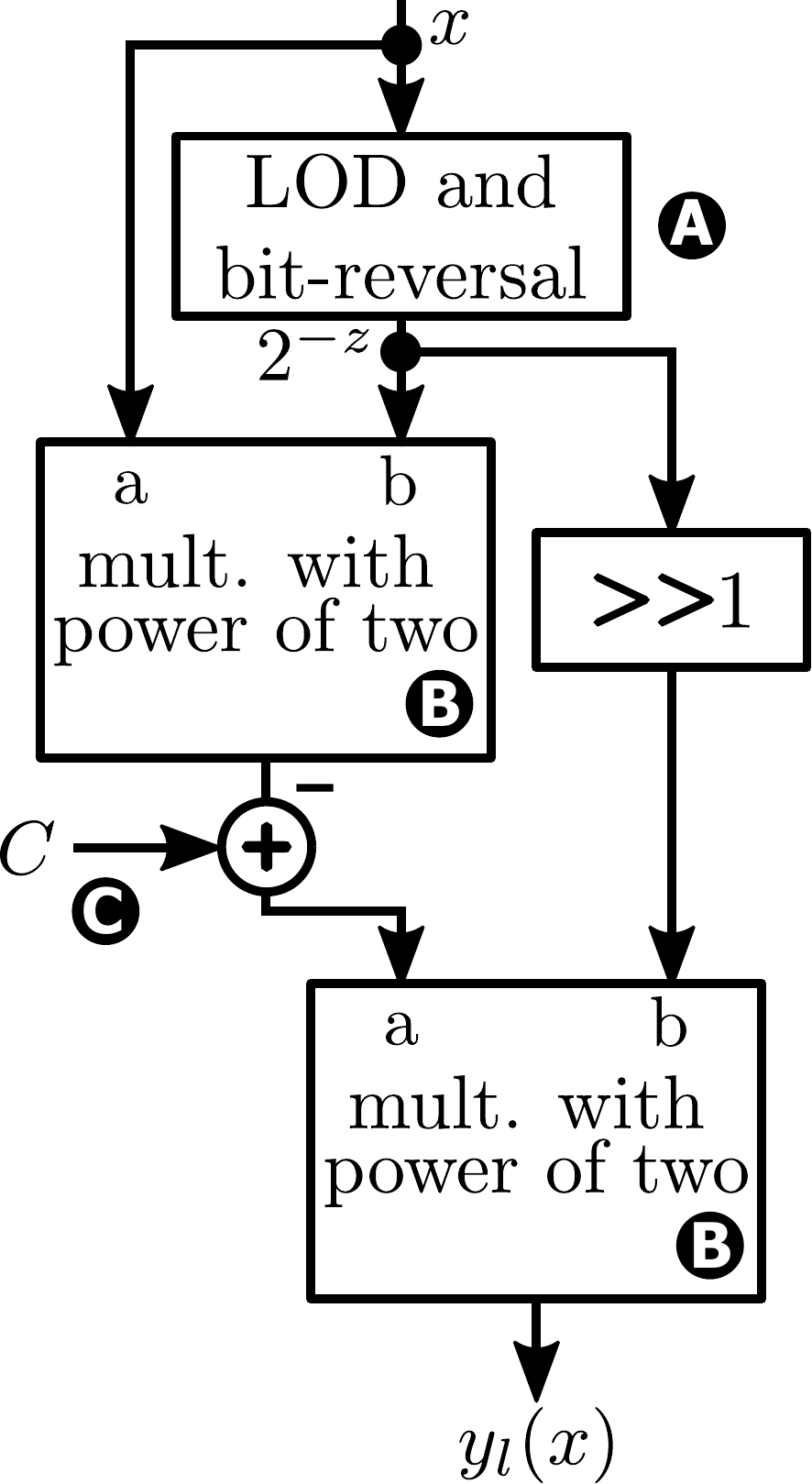}
	\caption{Approximate $1/x$ architecture.}
	\label{fig:OneOverArchitecture}
\end{figure}

The obtained approximation is especially interesting from a computational complexity point of view. Fig.~\ref{fig:OneOverArchitecture} shows an architecture for performing the calculation of (\ref{eqn:interpol}).
The architecture consists of a leading one detector (LOD) and bit-reversal block to calculate $2^{-z}$, two blocks performing a multiplication with a power of two, a constant shift, and an adder to implement (\ref{eqn:interpol}).
In the next section, we will discuss the blocks of the architecture in more detail. The encircled letters in the figure correspond to the subsections of Sect.~\ref{sect:buildingblocks}. 


\section{Building Blocks of Approximation Architecture} %
\label{sect:buildingblocks}
\subsection{Leading One Detector (LOD) and bit-reversal}
\begin{figure}[h]
	\centering
	\includegraphics[width=0.4\columnwidth]{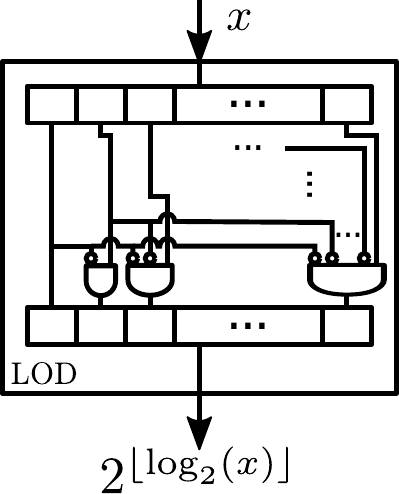}
	\caption{Leading One Detector.}
	\label{fig:LOD}
\end{figure}

To calculate $2^{z} = 2^{\lfloor \text{log}_2 (x) \rfloor}$, the architecture uses a so-called leading one detector\footnote{The name might be a little misleading, as the block does not only detect a leading one but also sets only the leading one in its output; we use this name following the wording that is common in the state-of-the-art literature as e.g. in \cite{ScalingLessNewton}.}. It can be performed by a structure as shown in Fig.~\ref{fig:LOD}. Its functionality is that it clears all bits below the leading one bit.  This is depicted in Fig.~\ref{fig:LOD} by the inverted inputs of the AND gates. For a bit in the output bit vector, all bits of the input vector that have a higher significance (i.e. are left of the output bit) are connected to inverted inputs of the corresponding AND gate.
Alternatively, the LOD can also be calculated using a bitwise AND operation of the bit-reversal of a number with its two complement \cite{ScalingLessNewton}. 

The value $2^{-z}$ is then obtained from $2^{z}$ by a simple bit-reversal (i.e. a mirroring of the bits at the radix point that is without cost in a hardware implementation).

\subsection{Multiplication with a power of two}
\begin{figure}[h]
	\centering
	\includegraphics[width=0.95\columnwidth]{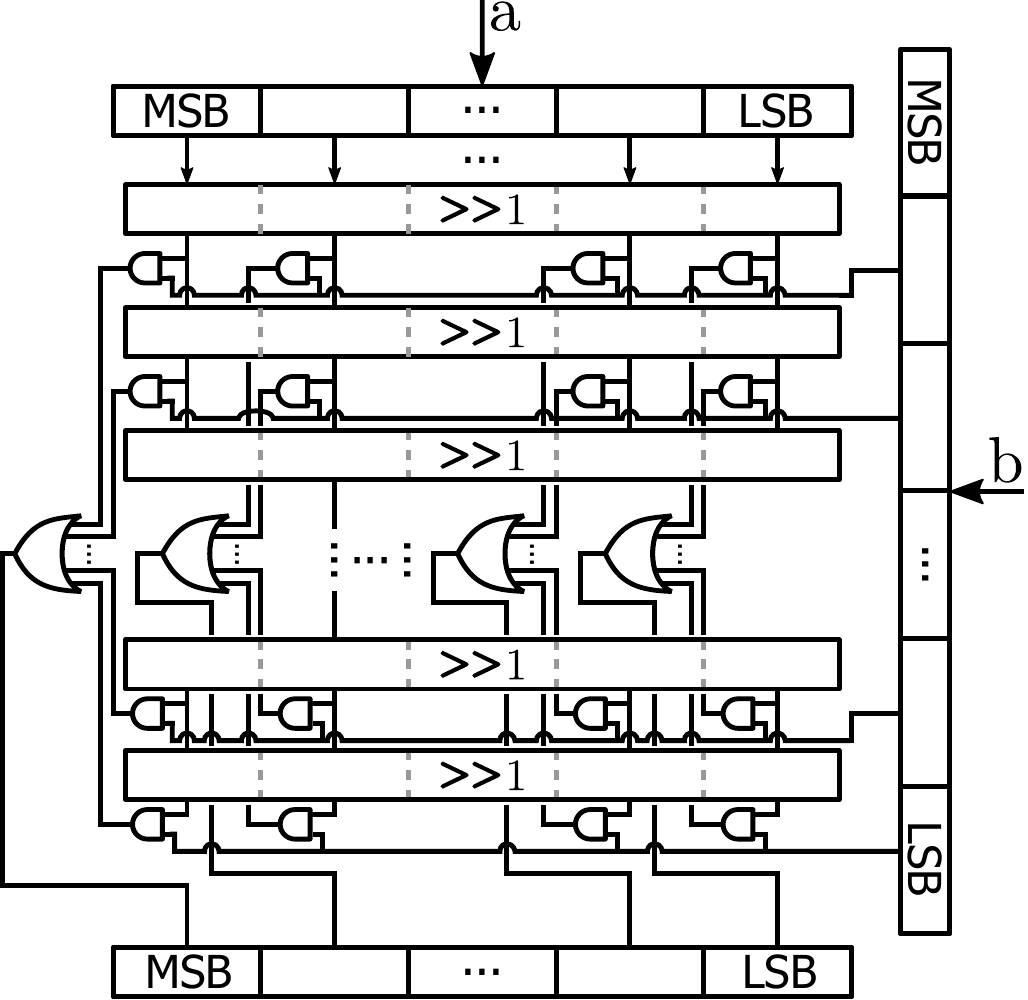}
	\caption{Multiplication with a power of two}
	\label{fig:OrShifter}
\end{figure}

The bit-reversed result of the LOD is required two times in (\ref{eqn:interpol}). One time in the multiplication with $2^{-z}$ and one time shifted right in the multiplication with $2^{-(z+1)}$. Both multiplication operations can either be realized with 
dedicated multipliers (in case they are available, e.g., on an FPGA) or with a multiplier-like structure having OR gates instead of full-adders as depicted (in its most simple) form in Fig.~\ref{fig:OrShifter}. For this, we assume that the power of two is always at the input $b$ of the multiplication block. In the figure, the blocks 
$>>1$
 arithmetically shift to the right by one. Because in the second operand, the power of two, only one bit can be set to one, OR operations between the $>>1$ layers are sufficient (instead of full adders when multiplying with an arbitrary number). 

\subsection{Subtraction from Constant}
The last part of the architecture is a subtraction from a constant $C$. 
The above derivation resulted in $C=3$. As we describe in the next section, this constant allows to further optimize this approach. Using a different constant $C$ changes the error behavior of the proposed method. As we will describe below, one can alternatively select this constant in a way such that the approximating curve is always below the optimal curve, or that the approximating curve has an average relative error of zero (compared to the exact solution of $1/x$). This allows optimizing the method for the desired error behavior.

\section{Optimizing the subtraction constant}
Due to the convexity of the $1/x$ function, a piecewise linearization using $C=3$ is always above the curve. This means that the error $e(x,C) = y_l(x,C)-1/x = y_l(x,3)-1/x$ is always positive. However, depending on the use-case for the approximate $1/x$ function, another error behavior might be desired. In this section, we discuss two alternatives: first, changing the approximate function such that it is always smaller or equal than the $1/x$ function (i.e. that the error $e(x)$ is always negative) and, second, changing the approximate function such that the average relative error is zero. As we show below, both alternatives can be obtained by just changing the factor $C$.

\subsection{An approximation that is always below $1/x$}
The idea behind this variant is to calculate the maximum error $e(x,3) = y_l(x,3)-1/x$ for every line segment and to subtract it from $y_l(x,3)$. This will lead to an approximation that is always below $1/x$. For this we first obtain the positions $x^*(z)$ of the maximum error 
of each linearization interval by deriving $\frac{d e(x,3)}{d x}$ and setting the derivative to zero:
\begin{align}
\frac{d\,e(x,3)}{d x} &= -2^{-z}2^{-(z+1)}+\frac{1}{x^2} = 0 \\
&\Rightarrow x^*(z) = 2^{\frac{2z+1}{2}}
\end{align}
Inserting these points $x^*(z)$ in $e(x,3)$ gives the values of the maximum error per line segment:
\begin{align}
(3-x^*(z) 2^{-z}) 2^{-(z+1)} - 1/x^*(z) = (3-2\sqrt{2}) 2^{-(z+1)}.
\end{align}
Subtracting this value from $y_l(x,3)$ for all linearization intervals ensures that the approximation curve is always below $1/x$ and results in 
\begin{align}
y_l(x,3)&-(3-2\sqrt{2}) 2^{-(z+1)} = \\
&= (3-x 2^{-z}) 2^{-(z+1)}-(3-2\sqrt{2}) 2^{-(z+1)}\\
&= (2\sqrt{2}-x 2^{-z}) 2^{-(z+1)} = y_l\left(x,2\sqrt{2} ) \right ).
\end{align}
This means that an approximation that is always below $1/x$ can be obtained by using the constant $C = 2\sqrt{2}$ in $y_l(x,C)$.

\subsection{Analytical description of relative error}
In general, the error of our reciprocal approximations can be described by numerical evaluations, but can also be described analytically. In our viewpoint, the error relative to the 
exact value
\begin{align}
r(x,C) = \frac{e(x,C)}{1/x} = x\,e(x,C)
\end{align}
is most interesting as it naturally puts the approximation error into perspective.

\begin{figure}[tb]
\begin{center}
\begin{tikzpicture}[spy using outlines={circle, magnification=2, connect spies}]
\begin{axis}[compat=newest, 
width=.9\columnwidth, height =.75\columnwidth,log basis y=10, grid, xlabel=Iteration, 
ylabel={  $r(x,3)$ in $\%$ }, 
xlabel={ $x$ }, 
xmin = 2,
xmax = 80,
ymax = 13,
legend style={at={(1.0,1.0)},anchor=north east, font=\scriptsize},
legend cell align=left,
legend columns = {1},
/pgf/number format/.cd, 1000 sep={}]
\addplot[color=blue, thick, style=solid] table[x index =0, y index =1] {./OneOverResultsErrors.dat};
\end{axis}
   in node[fill=white] at (magnifyglass);
\end{tikzpicture}
\caption{Relative approximation error over $x$ for $C=3$ \label{fig:errorC3}}
\end{center}
\end{figure}
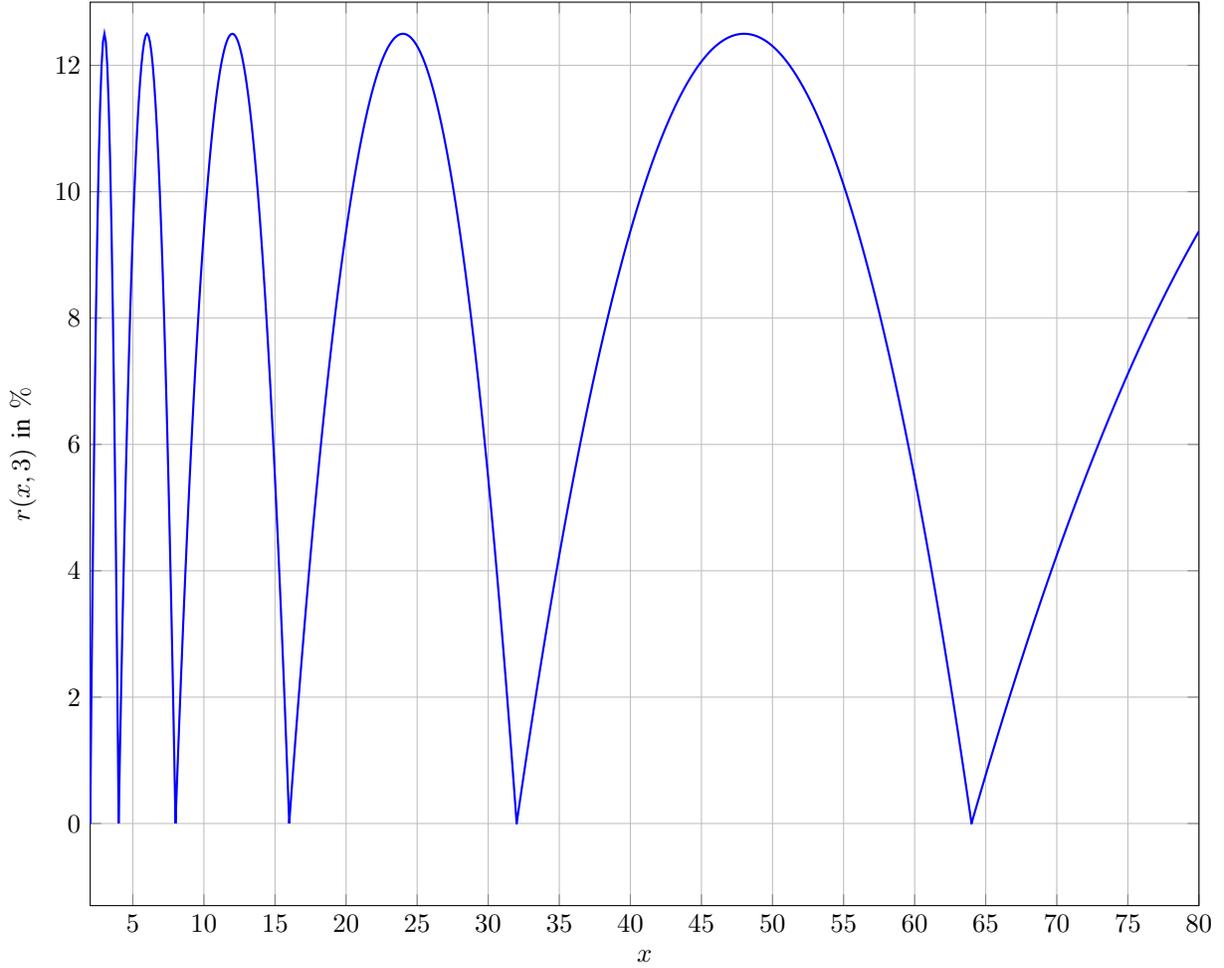

We analytically describe the average relative error (for an arbitrary constant $C$) by integrating over the error $r(x,C)$ and dividing the result by the interval width $2^z$. This results in the average approximation error for an arbitrary $C$ as
\begin{align}
r_\text{avg} (C) & = \frac{1}{2^z} \int_{2^z}^{2^z+1}  \frac{ (C-x2^{-z}) 2^{-(z+1)} - 1/x }{1/x} dx \nonumber\\
& = 3/4C-13/6. \label{eqn:avgrelerror}
\end{align}
For $C=3$ this results in a value for the average relative approximation error of $1/12 \approx 8.33\%$. Fig.~\ref{fig:errorC3} shows the relative error for $C=3$ plotted over $x$. For $C=2\sqrt{2}$ one obtains an average relative error $r_\text{avg} (2\sqrt{2}) \approx -4.53 \%$. 
Furthermore, (\ref{eqn:avgrelerror}) also gives insight on how to choose a constant resulting in 
\begin{align}
r_\text{avg} (C) = 0.
\end{align}
By solving this equation for $C$ one obtains $C=26/9 \approx 2.8889$ as a third variant of the proposed methods, achieving an average relative error of zero.

In the above derivation using $C=3$, we found the location of the maximum error point $x^*(z) = 2^{\frac{2z+1}{2}}$ as well as its value $(3-2\sqrt{2}) 2^{-(z+1)}$ for $C=3$. The position of the maximum relative error is slightly different, as one can obtain by deriving $r(x,C)$ with respect to $x$ 
\begin{align}
\frac{d\,r(x,C)}{d x} = \frac{d\,x\,e(x,C)}{d x} = C-x2^{-(z-1)}
\end{align}
Setting the derivative to zero gives the position of the extremal point
\begin{align}
C2^{(z-1)}.
\end{align}
Evaluating $r(x,C)$ at the position of its maximum gives for $C=3$
\begin{align}
r_\text{max}(3) = \frac{(3-2\sqrt{2}) 2^{-(z+1)}}{2^{-\frac{2z+1}{2}}} = \frac{3-2\sqrt{2}}{\sqrt{2}} = 12.5\%.
\end{align}
 
For the variants using $C=2\sqrt{2}$ and using $C=26/9$, the largest errors (in terms of absolute value) occur when approaching the interval borders $2^{z+1}$ from the left side. This means that the maximum error can be calculated at the value $x=2^{z+1}$ (more precisely at a infinitesimal small value before $2^{z+1}$ as the value $2^{z+1}$ already belongs to the next linearization interval) as 
\begin{align}
e(2^{z+1},C) = (C - 2^{z+1}2^{-z}) 2^{-(z+1)} - 2^{-(z+1)}.
\end{align}
Relating this value to $1/x=2^{-(z+1)}$ gives the relative error at the interval border, i.e. the maximum relative error for $C$ either $26/9$ or $2\sqrt{2}$ as
\begin{align}
r_\text{border}(C) &= \frac{e(2^{z+1},C)}{2^{-(z+1)}} \\
&= \frac{(C - 2) 2^{-(z+1)} - 2^{-(z+1)}}{2^{-(z+1)}} = C-3.
\label{eqn:largestborder}
\end{align}
For the values $C=2\sqrt{2}$ this gives the largest (in terms of its absolute value) relative error of $-17.16\%$. 
The largest (again seen from an absolute value perspective) relative error when using $C=26/9$ and (\ref{eqn:largestborder}) is $-11.1\%$. In Sect.~\ref{sect:MPE} we will show a simple method to significantly reduce the maximum relative error of both versions, when using $C=2\sqrt{2}$ or $C=26/9 \approx 2.8889$, respectively.

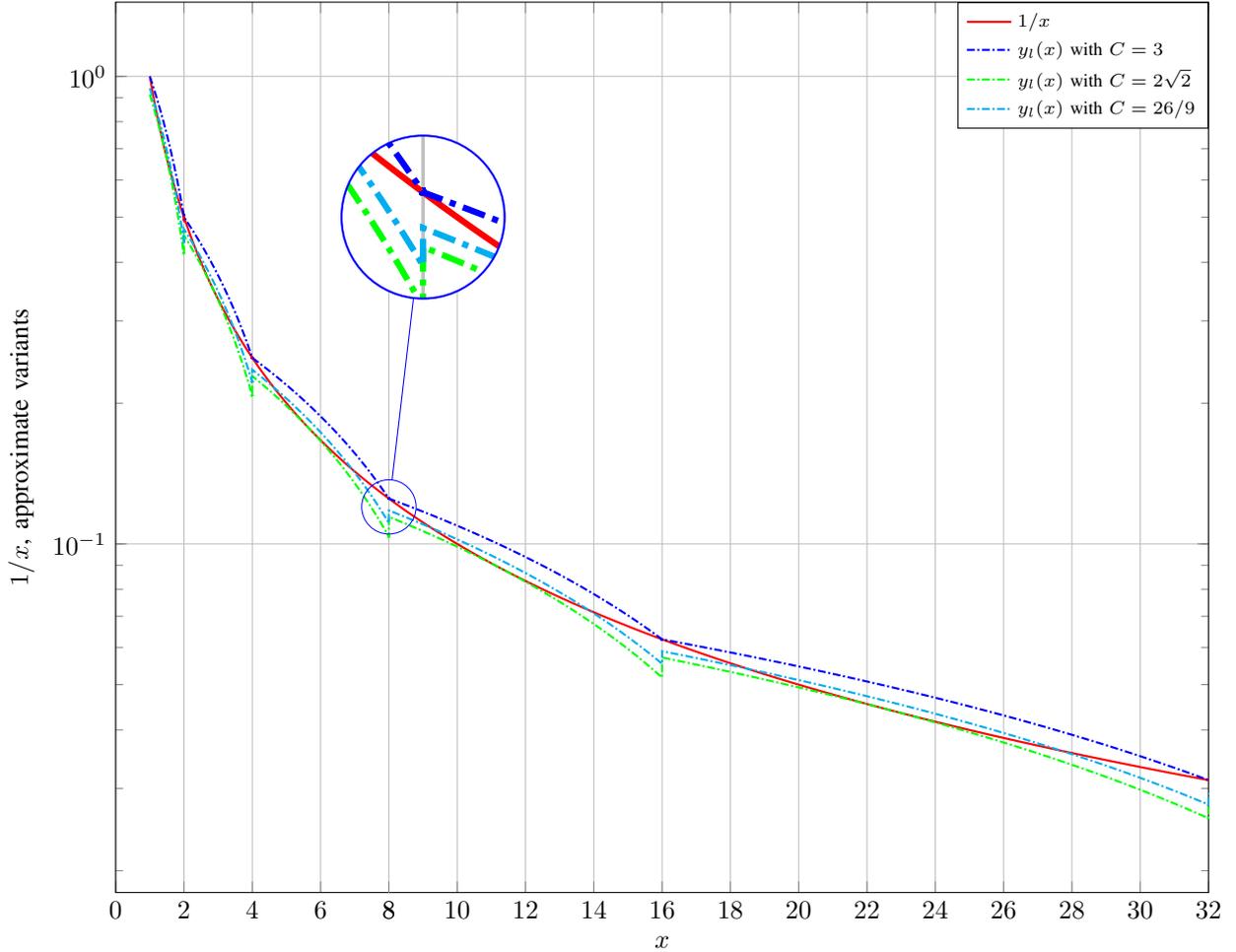
\begin{figure}[tb]
	\begin{center}
		\begin{tikzpicture}[spy using outlines={circle, magnification=3, connect spies}]
		\begin{semilogyaxis}[compat=newest, 
		width=.9\columnwidth, height =.75\columnwidth,log basis y=10, grid, xlabel=Iteration, 
		ylabel={  $1/x$, approximate variants }, 
		xlabel={ $x$ }, 
		xmin = 0,
		xmax = 32,
		extra x ticks={8},
		legend style={at={(1.0,1.0)},anchor=north east, font=\scriptsize},
		legend cell align=left,
		legend columns = {1},
		/pgf/number format/.cd, 1000 sep={}]
		\addplot[color=red,thick] table[x index =0, y index =1] {./OneOverResults.dat};
		\addlegendentry{ \scriptsize $1/x$ }
		\addplot[color=blue, thick, style=densely dashdotted] table[x index =0, y index =2] {./OneOverResults.dat};
		\addlegendentry{ \scriptsize $y_l(x)$ with $C=3$}
		\addplot[color=green, thick, style=densely dashdotted] table[x index =0, y index =3] {./OneOverResults.dat};
		\addlegendentry{ \scriptsize $y_l(x)$ with $C=2 \sqrt{2}$}
		\addplot[color=cyan, thick, style=densely dashdotted] table[x index =0, y index =4] {./OneOverResults.dat};
		\addlegendentry{ \scriptsize $y_l(x)$ with $C =  26/9$}
		\coordinate (spypoint) at (axis cs:8,.12);
		\coordinate (magnifyglass) at (axis cs:9,0.5);
		\end{semilogyaxis}
		\spy [blue, size=2.2cm] on (spypoint)
		in node[fill=white] at (magnifyglass);
		\end{tikzpicture}
		\caption{$1/x$ and approximation curves \label{fig:approxcurves}}
	\end{center}
\end{figure}

Fig.~\ref{fig:approxcurves} shows $1/x$ and the approximations with the different factors $C$. As one can observe from this figure, the curves show the characteristics we calculated $C$ for: the curve for $C=3$ is always above $1/x$, the curve for $C=2 \sqrt{2}$ is always below $1/x$ and the curve for 
$C =  26/9$  is in between. One can further notice that, for the last two variants ($C=2 \sqrt{2}$ and $C =  26/9$) due to the subtractions of the maximum and average errors, respectively, the approximations are not monotonic any longer. If monotonicity is required for an application we below propose a simple add-on not only ensuring monotonicity but also reducing the error as well.

\section{Implementation of approximate $1/x$}
\begin{figure}[h]
	\centering
	\includegraphics[width=0.55\columnwidth]{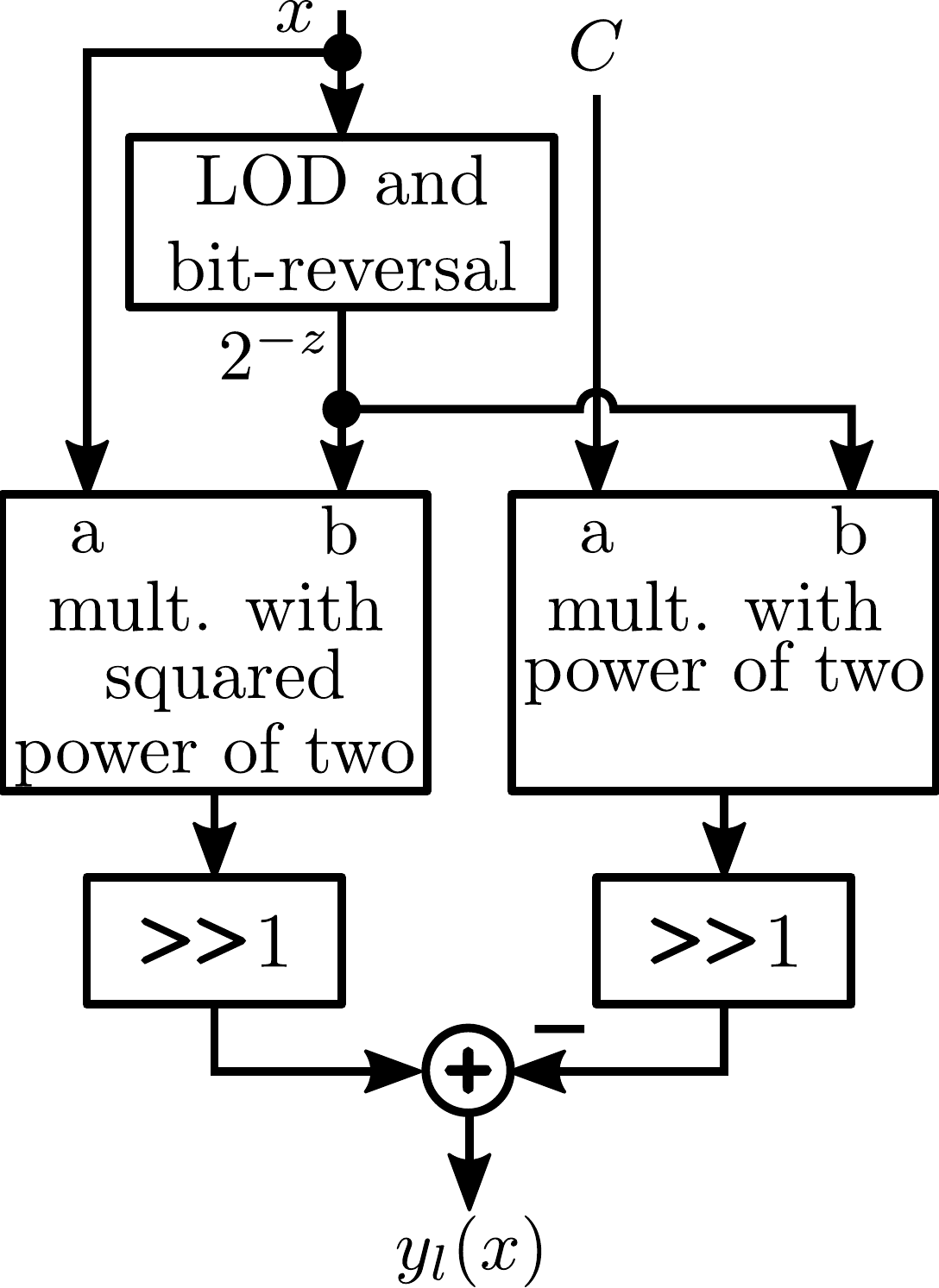}
	\caption{Approximate $1/x$ architecture - parallelized version.}
	\label{fig:OneOverArchitectureDouble}
\end{figure}

Although the whole structure of Fig.~\ref{fig:OneOverArchitecture} can be implemented using only combinatorics, one typically will introduce intermediate registers between the combinatorial sub-components. Such registers could be placed after the LOD block, after adding the constant $C$ and after the second multiplication with a power of two. This would allow for a high clock frequency while still requiring only $3$ clock cycles for calculating the approximate $1/x$ function. However, the structure can be re-arranged as shown in Fig.~\ref{fig:OneOverArchitectureDouble}. This structure originates from multiplying out the brackets of (\ref{eqn:interpol}). The changed structure requires a block for a multiplication with a squared power of two. Such a block can be easily built by altering the structure of Fig.~\ref{fig:OrShifter} such that each $>>1$ layer is replaced by $>>2$.
The architecture shown in Fig.~\ref{fig:OneOverArchitectureDouble} allows running both blocks, the one performing a multiplication with a power of two and the one performing a multiplication with a squared power of two, in parallel. A natural register placement for such a structure would be again after the LOD block and then after the final adder. This would reduce the number of required clock cycles to two. One can however implement both structures shown in Fig.~\ref{fig:OneOverArchitecture} and 
Fig.~\ref{fig:OneOverArchitectureDouble} without intermediate registers, then requiring only a single clock cycle for processing.
 In Sect.~\ref{sect:synthres}, we compare the different variants in terms of resource allocation and clock speed, with and without intermediate registers.

\subsection{Monotonicity preserving extension (MPE)}
\label{sect:MPE}
\begin{figure}[h]
	\centering
	\includegraphics[width=0.55\columnwidth]{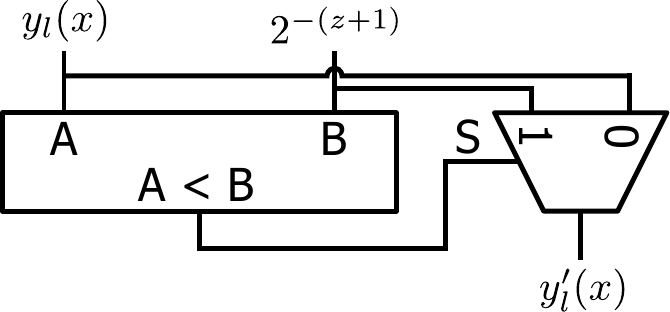}
	\caption{Monotonicity preserving extension (MPE).}
	\label{fig:monoaddon}
\end{figure}

As described above, when using the two discussed $C$-values smaller than $3$, the monotonicity of the approximation is no longer preserved. The error of 
the approximation is largest at the interval borders of the piecewise linearization. In this section, we describe a simple extension addressing both issues, preserving the monotonicity as well as reducing the maximum error of the approximation. As can be seen from Fig.~\ref{fig:approxcurves}, the error increases due to the fact that the approximation goes below the value $1/x$ of the interpolation interval border, i.e. $2^{-(z+1)}$. The idea of this extension is to limit the approximation value to be always higher or equal to $2^{-(z+1)}$. This can be performed by a comparison and a multiplexer, as it is schematically shown in Fig.~\ref{fig:monoaddon} (this architecture can be even more simplified by logic optimization).
This extension can be added to both the design of Fig.~\ref{fig:OneOverArchitectureDouble} and the design of Fig.~\ref{fig:OneOverArchitecture}, respectively, as both
allow to access the value $2^{-(z+1)}$ (the value of $2^{-z}$ shifted right by one). The extension does not only 
ensure the monotonicity of the approximation, but it also reduces the maximum error for $C$ values smaller than $3$ (for $C=3$ the extension would have no effect and is therefore redundant). This can be seen by calculating the point where the approximation is $2^{-(z+1)}$ (this is the first value that will be clipped by the architectural extension; as $1/x$ is monotonically decreasing, this is the value with the largest error of all clipped values in the linearization interval). Calculating this point can be performed by expressing
\begin{align}
(C-x 2^{-z}) 2^{-(z+1)} = 2^{-(z+1)} \\
\Rightarrow x_c(C) = (C-1) 2^{z}
\end{align}
leading to the first clipped point $x_c(C)$ in dependence of the value $C$.
Inputting the value of the approximation (with extension) $2^{-(z+1)}$ point in the error $e(x)$ gives

Relating this error to the correct value $\frac{1}{(C-1) 2^{z}}$ gives the relative error
\begin{align}
r_c(C) = \frac{3-C}{2}
\end{align}
at the first clipping point (one can also see from this derivation that for $C=3$ the clipping is redundant, as the clipping point would be at the border of the linearization interval, giving an error of zero).
For $C=2\sqrt{2}$ this gives $r_c(2\sqrt{2}) \approx 8.58\%$ and for $C =  26/9$ this gives $r_c(26/9) \approx 5.69\%$.

Fig.~\ref{fig:rel288} and Fig.~\ref{fig:relsqrt} shows the errors over multiple linearization intervals for the approximation with and without MPE for $C=2\sqrt{2}$ and $C =  26/9$, respectively. As one can see from these figures, the simple extension allows to significantly reducing the relative error. It leads to the smallest maximum relative deviation of $r_c(26/9)\approx 5.69 \%$ for $C =  26/9$ and to the smallest error range of $r_c(2\sqrt{2})\approx 8.58 \%$ for $C=2\sqrt{2}$ (the error for this approximation is always negative, thus limiting the negative error automatically limits the error range here).
\begin{figure}[tb]
\begin{center}
\begin{tikzpicture}[spy using outlines={circle, magnification=2, connect spies}]
\begin{axis}[compat=newest, 
width=.9\columnwidth, height =.75\columnwidth,log basis y=10, grid, xlabel=Iteration, 
ylabel={  $r(x, 26/9 )$ in $\%$ }, 
xlabel={ $x$ }, 
xmin = 2,
xmax = 80,
ymax = 7,
legend style={at={(1.0,1.0)},anchor=north east, font=\scriptsize},
legend cell align=left,
legend columns = {1},
/pgf/number format/.cd, 1000 sep={}]
\addplot[color=blue, thick, style=solid] table[x index =0, y index =2] {./OneOverResultsErrors.dat};
\addlegendentry{ \scriptsize $r(x, 26/9 )$}
\addplot[color=magenta, thick, style=densely dashdotted] table[x index =0, y index =2] {./OneOverResultsErrorsMPE.dat};
\addlegendentry{ \scriptsize $r(x, 26/9 )$ with MPE}
\end{axis}
   in node[fill=white] at (magnifyglass);
\end{tikzpicture}
\caption{Relative errors for $C = 26/9 $ with and without MPE \label{fig:rel288}}

\begin{tikzpicture}[spy using outlines={circle, magnification=2, connect spies}]
\begin{axis}[compat=newest, 
width=.9\columnwidth, height =.75\columnwidth,log basis y=10, grid, xlabel=Iteration, 
ylabel={  $r(x,2 \sqrt{2})$ in $\%$  }, 
xlabel={ $x$ }, 
xmin = 2,
xmax = 80,
ymax = 7,
legend style={at={(1.0,1.0)},anchor=north east, font=\scriptsize},
legend cell align=left,
legend columns = {1},
/pgf/number format/.cd, 1000 sep={}]
\addplot[color=blue, thick, style=solid] table[x index =0, y index =3] {./OneOverResultsErrors.dat};
\addlegendentry{ \scriptsize $r(x,2 \sqrt{2})$}
\addplot[color=magenta, thick, style=densely dashdotted] table[x index =0, y index =3] {./OneOverResultsErrorsMPE.dat};
\addlegendentry{ \scriptsize $r(x,2 \sqrt{2})$ and MPE}
\end{axis}
   in node[fill=white] at (magnifyglass);
\end{tikzpicture}
\caption{Relative errors for $C = 2 \sqrt{2} $ with and without MPE \label{fig:relsqrt}}

\begin{tikzpicture}[spy using outlines={circle, magnification=3, connect spies}]
\begin{semilogyaxis}[compat=newest, 
width=.9\columnwidth, height =.75\columnwidth,log basis y=10, grid, xlabel=Iteration, 
ylabel={  $1/x$, approximate variants }, 
xlabel={ $x$ }, 
xmin = 2,
xmax = 64,
legend style={at={(1.0,1.0)},anchor=north east, font=\scriptsize},
legend cell align=left,
legend columns = {1},
/pgf/number format/.cd, 1000 sep={}]
\addplot[color=red,thick] table[x index =0, y index =1] {./OneOverResults.dat};
\addlegendentry{ \scriptsize $1/x$ }
\addplot[color=blue, thick, style=densely dashdotted] table[x index =0, y index =2] {./OneOverResults.dat};
\addlegendentry{ \scriptsize $y_l(x)$ with $C=3$}
\addplot[color=green, thick, style=densely dashdotted] table[x index =0, y index =1] {./MonoResultsC2s2.dat};
\addlegendentry{ \scriptsize $y_l(x)$ with $C=2 \sqrt{2}$ and MPE}
\addplot[color=cyan, thick, style=densely dashdotted] table[x index =0, y index =1] {./MonoResultsC2.88.dat};
\addlegendentry{ \scriptsize $y_l(x)$ with $C =  26/9$ and MPE}
	\coordinate (spypoint) at (axis cs:16,.06);
  \coordinate (magnifyglass) at (axis cs:50,0.06);
\end{semilogyaxis}
\end{tikzpicture}
\caption{top: $1/x$ and approximation curves with MPE \label{fig:approxcurvesMPE}}
\end{center}
\end{figure}

\section{Synthesis Results}
\label{sect:synthres}
We synthesized the designs presented in this work for two FPGAs, a Cyclone V FPGA and a Startix V FPGA. 
We used Quartus Prime 17.1 for synthesis. For the presented synthesis results, version I of the architecture refers to the design schematically presented in Fig~\ref{fig:OneOverArchitecture}. We synthesized two variants of version I, one using the FPGAs onboard multipliers and one variant using the multiplication components of Fig.~\ref{fig:OrShifter}. Version II of the architecture refers to the design schematically presented in Fig.~\ref{fig:OneOverArchitectureDouble}. We synthesized all designs for $x$ values using an input bitwidth $B=16$, outputting a value with the same bitwidth. For the synthesis results presented in the 
Tab.~\ref{tab:cyclone} and Tab.~\ref{tab:stratix}, the columns a represent the synthesis results with the synthesis optimization mode set to ``Balanced'' but where register merging and register duplication was not allowed, respectively. The columns b represent synthesis results where the optimization mode was set to ``Performance (Aggressive)''. For this mode register merging and register duplication were allowed.
The tables contain the numbers of occupied adaptive logic modules (ALMs), the number of occupied flip-flops (output as ``Registers'') as well as the number of digital signal processing blocks (DSP blocks) that contain the dedicated multipliers (as well as registers and adder circuitry). All synthesis results include a registering of the input $\bf x$, to obtain valid timing results of the synthesis. In an application using the proposed architecture, such registers will typically be part of the architecture anyway as it is common to use registers to reduce the critical path. They are often used at the interfaces of sub-blocks of an architecture. In such a case, the number of registers of the synthesis tables for the presented architectures can be reduced by $16$.

The synthesis results for the Cyclone V are shown in Tab.~\ref{tab:cyclone}. 
The table shows a resource utilization below $1\%$ for each of the synthesized designs. Version I with full multipliers is the slowest but requires the smallest number of adaptive logic modules (ALMs) and Registers. However, it requires two DSP blocks while the other versions can be implemented without any DSP blocks. Especially version II shows very good results, combining low resource allocation with a maximum clock frequency up to nearly $200$MHz. According to the authors' experience such a clock frequency is seldom achieved for practical designs using this FPGA. Thus, the presented architecture will typically not be the limiting factor for the maximum clock frequency in typical designs. 
Version II is also favorable over the other synthesized designs because it only requires one clock cycle less than the other two designs. Including the MPE barely affects the clock frequency of the designs and only moderately increased the resource requirements of the design. However, in these synthesized variants it requires one clock cycle more to keep the high clock frequency. 
The synthesis results for the Stratix V FPGA are shown in Tab.~\ref{tab:stratix}. For these results, the identical VHDL codes have been used as for the results of Tab.~\ref{tab:stratix}. As expected, the clock frequency of the synthesized design increases for this FPGA. For this FPGA the maximum clock frequency was achieved for version II of the design, yielding up to about $380$MHz. As expected because of a similar structure of the FPGA's building blocks (e.g the ALMs) the resource allocations in terms of the allocated numbers have been comparable between the Stratix V and the Cyclon V. The main differences are the different maximum clock frequencies.
\begin{table}[ht]
\centering
\caption{Synthesis results for different implementations on {Cyclone V: 5CSXFC6D6F31C6} \label{tab:cyclone}}
\setlength\tabcolsep{3pt} 
\setlength\extrarowheight{6pt}
\begin{tabular}{|p{2.7cm}|p{0.75cm}|p{.75cm}|p{.75cm}|p{.75cm}|p{.75cm}|p{.75cm}|}
	\hline
	                         &                                                                         \multicolumn{6} {c|}{Version}                                                                         \\
	                         & \multicolumn{2} {p{1.5cm}|}{I with full multipliers} & \multicolumn{2} {p{1.5cm}|}{I with Fig.~\ref{fig:OrShifter} components} & \multicolumn{2} {p{1.5cm}|}{\centering II}   \\
	                         & \centering a        & \centering b                   & \centering a & \centering b                                             & \centering a & \centering \arraybackslash b \\ \hline
	ALMs (of $41,910$)       & $38$                & $33$                           & $127$        & $208$                                                    & $89$         & $125 $                       \\
	Registers (of 166\;036)  & $36 \footnote[3]{}$ & $48 \footnote[3]{}$            & $88$         & $196$                                                    & $54$         & $107$                        \\
	DSP blocks (of 112)      & $2$                 & $2$                            & $0$          & $0$                                                      & $0$          & $0$                          \\
	Fmax slow 1.1V 85C (MHz) & $134.7$             & $151.33$                       & $155.84$     & $199.36$                                                 & $142.92$     & $199.04$                     \\ \hline
	Clock cycles             & \multicolumn{2} {p{1.5cm}|}{\centering 3}            & \multicolumn{2} {p{1.5cm}|}{\centering 3}                               & \multicolumn{2} {p{1.5cm}|}{\centering  2}  \\ \hline
	                         &                                                                \multicolumn{6} {c|}{Version (including MPE)}                                                                 \\ \hline
	ALMs (of $41,910$)       & $49$                & $51 $                          & $135$        & $222$                                                    & $101$        & $143$                        \\
	Registers (of 166\;036)  & $66$\footnote[3]{}                & $96 \footnote[3]{}$            & $108$        & $219$                                                    & $74$         & $166$                        \\
	DSP blocks (of 112)      & $2$                 & $2$                            & $0$          & $0$                                                      & $0$          & $0$                          \\
	Fmax slow 1.1V 85C (MHz) & $127.32$            & $153.89$                       & $153.09$     & $194.51$                                                 & $142.39$     & $211.46$                     \\ \hline
	Clock cycles             & \multicolumn{2} {p{1.5cm}|}{\centering 4}            & \multicolumn{2} {p{1.5cm}|}{ \centering  4}                             & \multicolumn{2} {p{1.5cm}|}{ \centering 3}  \\ \hline
\end{tabular}
\begin{flushleft}
$^{\textrm{3}}$: the residual registers have been used from the DSP Blocks.
\end{flushleft}
\end{table}

\begin{table}[ht]
\centering
\caption{Synthesis results for different implementations on {Stratix V: 5SGSMD5K2F40C2} \label{tab:stratix}}
\setlength\tabcolsep{3pt} 
\setlength\extrarowheight{6pt}
\begin{tabular}{|p{2.7cm}|p{0.75cm}|p{.75cm}|p{.75cm}|p{.75cm}|p{.75cm}|p{.75cm}|}
	\hline
	                          &                                                                         \multicolumn{6} {c|}{Version}                                                                         \\
	                          & \multicolumn{2} {p{1.5cm}|}{I with full multipliers} & \multicolumn{2} {p{1.5cm}|}{I with Fig.~\ref{fig:OrShifter} components} & \multicolumn{2} {p{1.5cm}|}{\centering II}   \\
	                          & \centering a & \centering b                          & \centering a & \centering b                                             & \centering a & \centering \arraybackslash b \\ \hline 
	ALMs (of $172,600$)       & $42$         & $39 $                                 & $125$        & $208$                                                    & $89$         & $126$                        \\
	Registers (of $706,560$)  & $36$\footnote[4]{}         & $51$\footnote[4]{}                                 & $88$         & $179$                                                    & $54$         & $114$                        \\
	DSP blocks (of $1,590$)   & $2$          & $2$                                   & $0$          & $0$                                                      & $0$          & $0$                          \\
	Fmax Slow 900mV 85C (MHz) & $247.89$     & $252.33$                              & $296.74$     & $368.42$                                                 & $282.49$     & $387.6$                      \\ \hline
	Clock cycles              & \multicolumn{2} {p{1.5cm}|}{\centering 3}            & \multicolumn{2} {p{1.5cm}|}{\centering 3}                               & \multicolumn{2} {p{1.5cm}|}{\centering 2}   \\ \hline
	                          &                                                                \multicolumn{6} {c|}{Version (including MPE)}                                                                 \\ \hline
	ALMs (of $172,600$)       & $53$         & $52$                                  & $135$        & $225$                                                    & $101$        & $142$                        \\
	Registers (of $706,560$)  & $66$\footnote[4]{}         & $94$\footnote[4]{}                                  & $108$        & $220$                                                    & $74$         & $187$                        \\
	DSP blocks (of $1,590$)   & $2$          & $2$                                   & $0$          & $0$                                                      & $0$          & $0$                          \\
	Fmax Slow 900mV 85C (MHz) & $249.38$     & $248.94$                              & $309.6$      & $364.83$                                                 & $274.5$      & $388.5$                      \\ \hline
	Clock cycles              & \multicolumn{2} {p{1.5cm}|}{\centering 4}            & \multicolumn{2} {p{1.5cm}|}{\centering 4}                               & \multicolumn{2} {p{1.5cm}|}{\centering 3}   \\ \hline
\end{tabular}
\begin{flushleft}
$^{\textrm{4}}$: the residual registers have been used from the DSP Blocks.
\end{flushleft}
\end{table}
Both tables show that all the presented variants require a low number of the resources of the respective FPGAs (signifcantly below $1\%$). As one can furthermore see, especially by register duplication the synthesis is able to increase the clock frequency (these registers are in parallel and thus do not increase the number of required clock cycles).

We also synthesized version II using an output register and no intermediate registers. The results thereof are shown in Tab.~\ref{tab:noreg}. One can see from these results that even when using no intermediate registers, on the Cyclone V, a clock frequency of about $100$MHz and above could be achieved. For the Stratix V, clock frequencies of above $200$MHz could be achieved. The ALM requirements where around $66$ without MPE and around $86$ with MPE. Please note that to achieve valid timing estimations from the synthesis, we again included a registering of the input in the design (accounting for $16$ flip-flops of the synthesis results; $16$ more were used for the output register and one additional to hold a valid strobe signal for the output). The implementations of Tab.~\ref{tab:noreg} allow performing the described approximations within a single clock cycle.

\begin{table}[ht]
	\centering
	\caption{Synthesis results for version II without intermediate registers on {Cyclone V: 5CSXFC6D6F31C6} and on {Stratix V: 5SGSMD5K2F40C2} \label{tab:noreg}}
	\setlength\tabcolsep{3pt} 
	\setlength\extrarowheight{6pt}
	\begin{tabular}{|p{2.7cm}|p{1.1cm}|p{1.1cm}|p{1.1cm}|p{1.1cm}|}
		\hline
		& \multicolumn{2}{p{2.2cm}|}{ {Cyclone V:} $\;\;\;\;\;$ 5CSXFC6D6F31C6} & \multicolumn{2}{p{2.2cm}|}{Stratix V: $\;\;\;\;\;$ 5SGSMD5K2F40C2} \\
		& \centering a & \centering b                                           & \centering a & \centering \arraybackslash b                        \\          \hline
		ALMs (of $172,600$)       & $66$         &  $66$                                                 & $67$         &    $65$                                            \\
		Registers (of $706,560$)  & $33$         &  $33$                                                 & $33$         &    $33$                                            \\
		DSP blocks (of $1,590$)   & $0$          &  $0$                                                  & $0$         &    $0$                                              \\
		Fmax Slow 900mV 85C (MHz) & $135.23$     &  $134.68$                                             & $262.26$   &    $268.31$                                         \\ \hline
		&                                                   \multicolumn{4} {c|}{(including MPE)}                                                    \\ \hline
		ALMs (of $172,600$)       & $84$       & $89$                                                   & $86$        &   $90$                                             \\
		Registers (of $706,560$)  & $33$       & $33$                                                   & $33$        &   $33$                                             \\
		DSP blocks (of $1,590$)   & $0$        & $0$                                                    & $0$        &   $0$                                               \\
		Fmax Slow 900mV 85C (MHz) & $99.8$     & $117.18$                                               & $213.27$        &   $240.56$                                          \\ \hline
	\end{tabular}
\end{table}

\section{Comparision with State-of-the-Art}
As described in the introduction of this work, the aim of most state-of-the-art algorithms is to approximate the reciprocal function with a high precision. Although a direct comparison to such algorithms might not be completely fair (due to the different design aims of the architectures as commented above), one can use synthesis results of state-of-the-art works to put this work into perspective. 
In Tab.~\ref{tab:compari} we collected three synthesis results of efficient reciprocal FPGA implementations reported in literature. 
\begin{table}[h]
	\centering
	\caption{ Literature reports on efficient reciprocal implementations \label{tab:compari}}
	\setlength\tabcolsep{3pt} 
\setlength\extrarowheight{6pt}
\begin{tabular}{|p{2.7cm}|p{1.5cm}|p{1.5cm}|p{1.5cm}|}
	\hline
	Work reported in        & [7]				& [7]			& [5]			\\
			\hline
	FPGA                    & VIRTEX-4 SX35     & VIRTEX-7 690T & Stratix-V 5SGXMA7 \\
	\hline
	Slices                  & 347               & not reported  & 339   \\
	LUT                     & 372               & 111           &       \\
	Registers               & 568               & 240           & 73    \\
	DSP blocks              & 7                 & 6             & 5     \\
	Clock cycles            & 25                & 25            & 3     \\
	Clock frequency         & 294.1             & 740           & 68.62 \\
	\hline
\end{tabular}
\end{table}

The results have all been synthesized for $16$ bit fixed point precision numbers as it was also done for synthesis results presented in this work. As one can see from this table, the reported methods require much higher hardware resources and also require higher computation times due to either lower clock frequencies of a larger number of required clock cycles for completing the calculation. This is due to the different aims of the cited works, providing a high precision implementation of the reciprocal function. If one requires such high precision, the cited designs efficiently allow calculating the reciprocal function. If the precision of our proposed approximate approach is sufficient for an application at hand, the architectures presented in this work require  significantly less hardware resources as well as significantly less computation time. 

\section{Application example I: Sparse Kaczmarz Algorithm}
In this section, we show the effect of using the approximate $1/x$ function in a sparse estimation setting using a sparse LMS filter based on linearized Bregman iterations \cite{OLBILMS, ISCAS2017}. 

\begin{figure}[ht]
	\begin{center}
		\includegraphics[width=\linewidth]{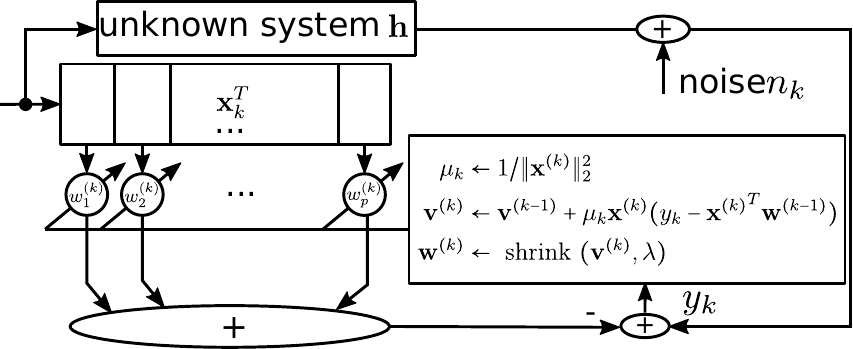}
		\caption{System identification with sparse LMS \label{fig:SparseLMS}}		
	\end{center}		 		
\end{figure}

The main structure of such a sparse LMS filter is shown in Fig.~\ref{fig:SparseLMS}. Here the sparse LMS filter is shown in a system identification scenario, estimating a system impulse response $\bf h$. The estimation results are the filter coefficients of the sparse LMS filter ${\bf w}^{(k)} = [{w}_1^{(k)}, {w}_2^{(k)}, \ldots, {w}_p^{(k)}]$.

\begin{figure}[hb]
\begin{center}
\begin{tikzpicture}[spy using outlines={circle, magnification=3, connect spies}]
\begin{semilogyaxis}[compat=newest, 
width=.9\columnwidth, height =.75\columnwidth,log basis y=10, grid, xlabel=Iteration, 
ylabel={  $\text{mean }\|{\bf w}^{(k)} - {\bf h}\|_2^2$ }, 
xlabel={ iteration $i$ }, 
legend style={at={(1.0,1.0)},anchor=south east, font=\scriptsize},
legend cell align=left,
legend columns = {1},
/pgf/number format/.cd, 1000 sep={}]
\addplot[color=red,thick] table[x index =0, y index =1] {./SLMSResult.dat};
\addlegendentry{ \scriptsize Sparse LMS using $1/x$ }
\addplot[color=blue, thick, style=densely dashed] table[x index =0, y index =2] {./SLMSResult.dat};
\addlegendentry{ \scriptsize Sparse LMS using  $y_l(x)$ w. $C\!=\!2\sqrt{2}$ }
\addplot[color=green, thick, style=densely dashed] table[x index =0, y index =3] {./SLMSResult.dat};
\addlegendentry{ \scriptsize Sparse LMS using  $2^{-\lceil log_2(x) \rceil}$}
\addplot[color=cyan, thick, style=densely dotted] table[x index =0, y index =4] {./SLMSResult.dat};
\addlegendentry{ \scriptsize Sparse LMS using  $y_l(x)$ w. $C\!=\!2\sqrt{2}$ and MPE }
 	\coordinate (spypoint) at (axis cs:410,.14);
   \coordinate (magnifyglass) at (axis cs:120,0.08);
\end{semilogyaxis}
 \spy [blue, size=1.5cm] on (spypoint)
    in node[fill=white] at (magnifyglass);
\end{tikzpicture}
\caption{Average squared error norm over iterations \label{fig:SLMSResult}}
\end{center}
\end{figure}
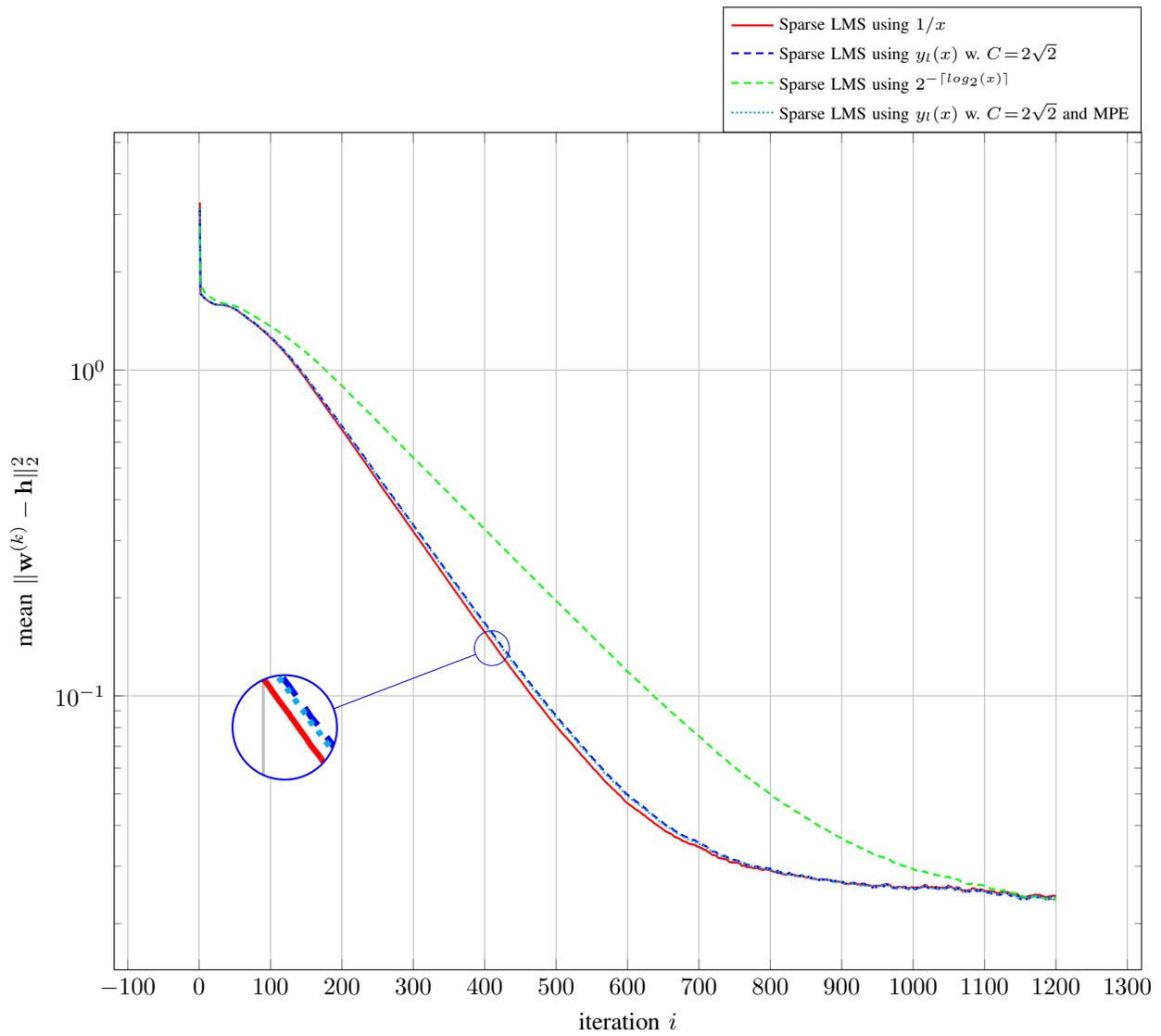

The figure shows the update equations for the filter coefficients ${\bf w}^{(k)}$. The first equation uses a reciprocal calculation for calculating the step size $\mu_k$. We simulated sparse system identification for system impulse responses ${\bf h}$ of length $30$ with $3$ non-zero elements.  
 Fig.~\ref{fig:SLMSResult} show an example result obtained for this sparse estimation example up to $1200$ sparse LMS iterations. The input signal of the sparse LMS was sampled from a uniform random distribution out of the interval $[0,1]$. Here, we show results obtained in double precision to prevent other quantization effects from influencing the results. To obtain the estimation results we used three different variants, one using the optimal $1/x$ function, one using the approximation of this work with $C=2\sqrt{2}$, with and without MPE, respectively, and one using 
 $2^{-\lceil log_2(x) \rceil}$. Because here the $1/x$ calculation is used for the step size of an iterative algorithm (where too large values can lead to divergence) the variant with $C=2 \sqrt{2}$ was used. The simpler approximation $2^{-\lceil log_2(x) \rceil}$ was used to provide a base-line for comparison. 
The results show the square root of the averaged squared error norm: $\sqrt{\text{mean }\|\hat{\bf w}_i - w\|}_2^2$ over the iterations $i$. The averaging was performed over $1000$ randomly selected test-cases. $\hat{\bf w}^{(k)}$ are the estimated filter coefficients at iteration $k$ and ${\bf h}$ are the true filter coefficients. As the results in Fig.~\ref{fig:SLMSResult} show, for this use-case, the approximation discussed in this work comes very close to the optimum solution while the simpler approximation $2^{-\lceil log_2(x) \rceil}$ leads to a significant increase of the number of required iterations for the Sparse LMS to converge. Because the results of the approximation with $C=2\sqrt{2}$ are already close to the optimum solution, for this test-case, MPE only provides slight performance gains. 

\section{Application example II: k-means Algorithm}
For the second application example, we investigated the use of our approximate $1/x$ function in the k-means clustering algorithm \cite{TibshiISL}. This algorithm iteratively performs the assignment of vectors into one of $K$ clusters. Each iteration consists of two steps
\begin{enumerate}
	\item for every cluster: calculate the cluster mean among all vectors that belong to a cluster
	\item (re-)assign each vector to the cluster with the closest mean 
\end{enumerate}
The first step of an iterations consists of a mean calculation that involves a $1/|C_k|$ operation (here, $|C_k|$ being the number of vectors belonging to cluster $C_k$; $C_k$ being the corresponding set of vectors).

\begin{figure}
	\begin{tikzpicture}
	\begin{axis}[%
	scatter/classes={%
		1={mark=square*,cyan,opacity=0.35,draw opacity=0},%
		2={mark=square*,red,opacity=0.35,draw opacity=0},%
		3={mark=square*,yellow,opacity=0.35,draw opacity=0},
		4={mark=square*,green,opacity=0.35,draw opacity=0},
		5={mark=o, line width=1pt,mark size=3pt,black},
		6={mark=triangle, line width=1pt,mark size=3pt,black},
		7={mark=diamond, line width=1pt,mark size=3pt,black}}, 	legend entries={ $\;\;$ optimal,
		$\;\;$ using approximate $1/x$ using $C=26/9$ and MPE, 
		$\;\;$ using $2^{-\lceil log_2(x) \rceil}$}, legend style={at={(1.0,1.0)},anchor=south east, font=\scriptsize},legend cell align={left}]
	\addplot[scatter,only marks,%
	scatter src=explicit symbolic, forget plot]%
	table[meta=label] {./kmeanssingle.dat};
	\addlegendimage{only marks, mark=o,line width=1pt,mark size=3pt}
	\addlegendimage{only marks, mark=triangle,line width=1pt,mark size=3pt}
	\addlegendimage{only marks, mark=diamond,line width=1pt,mark size=3pt}
	\end{axis}
	\end{tikzpicture}
	\caption{Example results of k-means application. \label{fig:kmeanssingle}}
\end{figure}
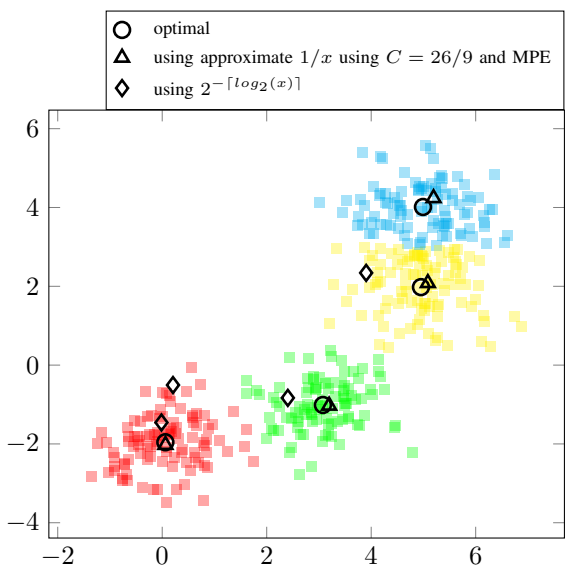

We compared the results using the optimal $1/x$ function for the mean with the results using our approximation with $C= 26/9$ and MPE. For this, we simulated $K$ clusters by randomly placing center point on the $2$-dimensional plain where for each coordinate a random integer out of $\{-4,-3,-2,-1,0,1,2,3,4\}$ was chosen. Around each center point we randomly placed $100$ points drawn from a Gaussian distribution with a covariance matrix 
\begin{align}
\begin{pmatrix}
0.5  & 0.05\\
0.05 & 0.5
\end{pmatrix}.
\end{align}
Fig.~\ref{fig:kmeanssingle} shows an example result showing the clustering obtained by the k-means algorithm. The coloring shows the membership to a cluster (here we use the coloring from the optimal result as the cluster membership might differ depending on the used approach for calculating/approximating the $1/x$ function). The circles mark the cluster centers 
using the optimal $1/x$ function. Triangles mark the cluster centers using the approximate variants. For comparison, we again plotted the results when using $2^{-\lceil log_2(x) \rceil}$ as a coarse approximation for the $1/x$ function. The corresponding cluster centers are plotted as diamonds in 
Fig.~\ref{fig:kmeanssingle}. For all three approaches, we used the same random initialization points for the start centers at the beginning of the algorithm.

This figure shows that using the very coarse approximation $2^{-\lceil log_2(x) \rceil}$ resulting in large deviations. Contrary, with the proposed approximation, the deviation is only minor compared to the optimal solution.

We furthermore compared the performance of $k$-means with these three methods to calculate/approximate the $1/x$ function in a more detailed simulation study. For performance comparison we used the sum of squared distances (SSD) of each vector ${{\bf v}_i = [v_{i1},v_{i2}, \ldots v_{ip}]}$ to the center 
of its cluster $C_k$:
\begin{align}
\text{SSD} = \sum_{k=1}^{K} \sum_{i: {\bf v}_i \in C_k} \| {\bf v}_i - \bar{\bf v}_k\|_2^2.
\label{eqn:SSD}
\end{align}
With the cluster centers
\begin{align}
\bar{\bf v}_k=\frac{1}{|C_k|} \sum_{i: {\bf v}_i \in C_k} {\bf v}_i.
\end{align}
This metric is minimized by the k-means algorithm (at least up to a local minimum), so it provides a natural quality measure for the results using the approximations for $1/x$.

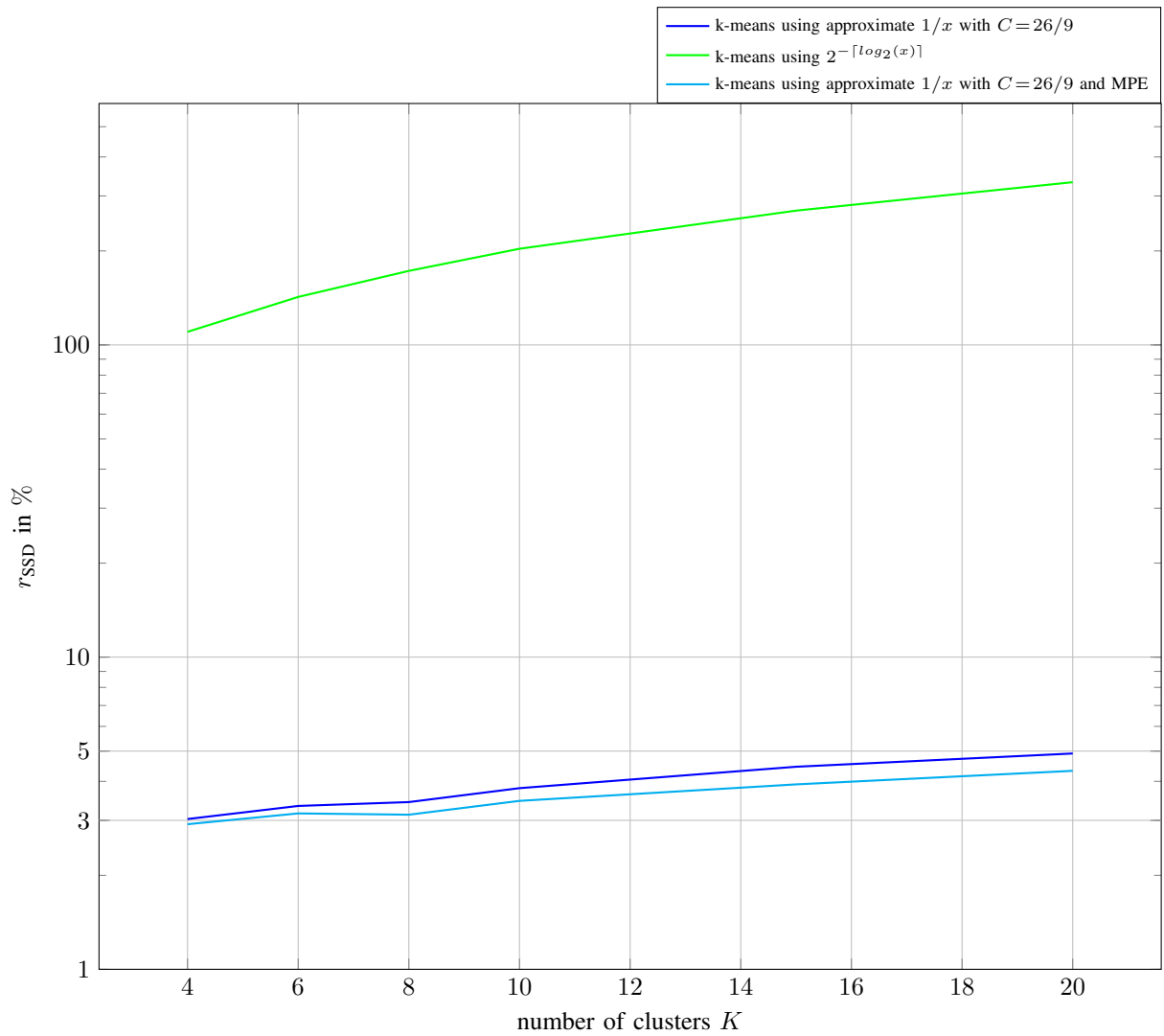
\begin{figure}[t]
	\begin{center}
		\begin{tikzpicture}
		\begin{semilogyaxis}[compat=newest, 
		width=.9\columnwidth, height =.75\columnwidth,log basis y=10, grid, xlabel=Iteration, 
		ylabel={  $r_{\text{SSD}}$ in $\%$ }, 
		xlabel={ number of clusters $K$ }, 
		ymin = 1,
		log ticks with fixed point,
		y tick label style={/pgf/number format/1000 sep=\,},
		extra y ticks={3,5},
		legend style={at={(1.0,1.0)},anchor=south east, font=\scriptsize},
		legend cell align=left,
		legend columns = {1},
		/pgf/number format/.cd, 1000 sep={}]
		\addplot[color=blue,thick] table[x=noclusters, y=AvgDiffCostApproxPerc] {./KMeans_norm2_10000_2.888889.txt};
		\addlegendentry{ \scriptsize k-means using approximate $1/x$ with $C\!=\!26/9$ }
		\addplot[color=green,thick] table[x=noclusters, y=AvgDiffCostApproxPow2Perc] {./KMeans_norm2_10000_2.888889.txt};
		\addlegendentry{ \scriptsize k-means using $2^{-\lceil log_2(x) \rceil}$ }
				\addplot[color=cyan,thick] table[x=noclusters, y=AvgDiffCostApproxPerc] {./KMeans_norm2_10000_2.888889MPE.txt};
		\addlegendentry{ \scriptsize k-means using approximate $1/x$ with $C\!=\!26/9$ and MPE}		
		\end{semilogyaxis}
		\end{tikzpicture}
		\caption{Relative increase of costs for k-means when using approximate $1/x$ functions \label{fig:approxkmeans}}
	\end{center}
\end{figure}

For this we plotted the relative increase of the $\text{SSD}$-value caused by an approximation in percent:
\begin{align}
r_{\text{SSD}} = (\text{SSD}_{approx} - \text{SSD})/\text{SSD} * 100\%,
\end{align}
with $\text{SSD}$ as the value of (\ref{eqn:SSD}) when using the exact value of $1/x$ and with 
$\text{SSD}_{approx}$ as the value of (\ref{eqn:SSD}) when using an approximation of $1/x$. Fig.~\ref{fig:approxkmeans} shows the results for different k-means clustering scenarios using different numbers of $K$. For each number of $K$, the results have been averaged over $10000$ test cases. We again used the coarse approximation $2^{-\lceil log_2(x) \rceil}$ for $1/x$ for performance comparison. As one can see in Fig.~\ref{fig:approxkmeans}, the performance when using our proposed approximation with $C=26/9$ is only $3-5\%$ worse than the ideal solution. When using MPE the performance could be further improved for most $K$ values by about $0.5\%$ on average. The results when using $2^{-\lceil log_2(x) \rceil}$ are on average more than $100\%$ worse than the optimal ones, leading to, especially for a large number of clusters, unusable results.

\section{Application example III: Neural Network regression}

\begin{figure}[t]
	\begin{center}
		\includegraphics[width=\linewidth]{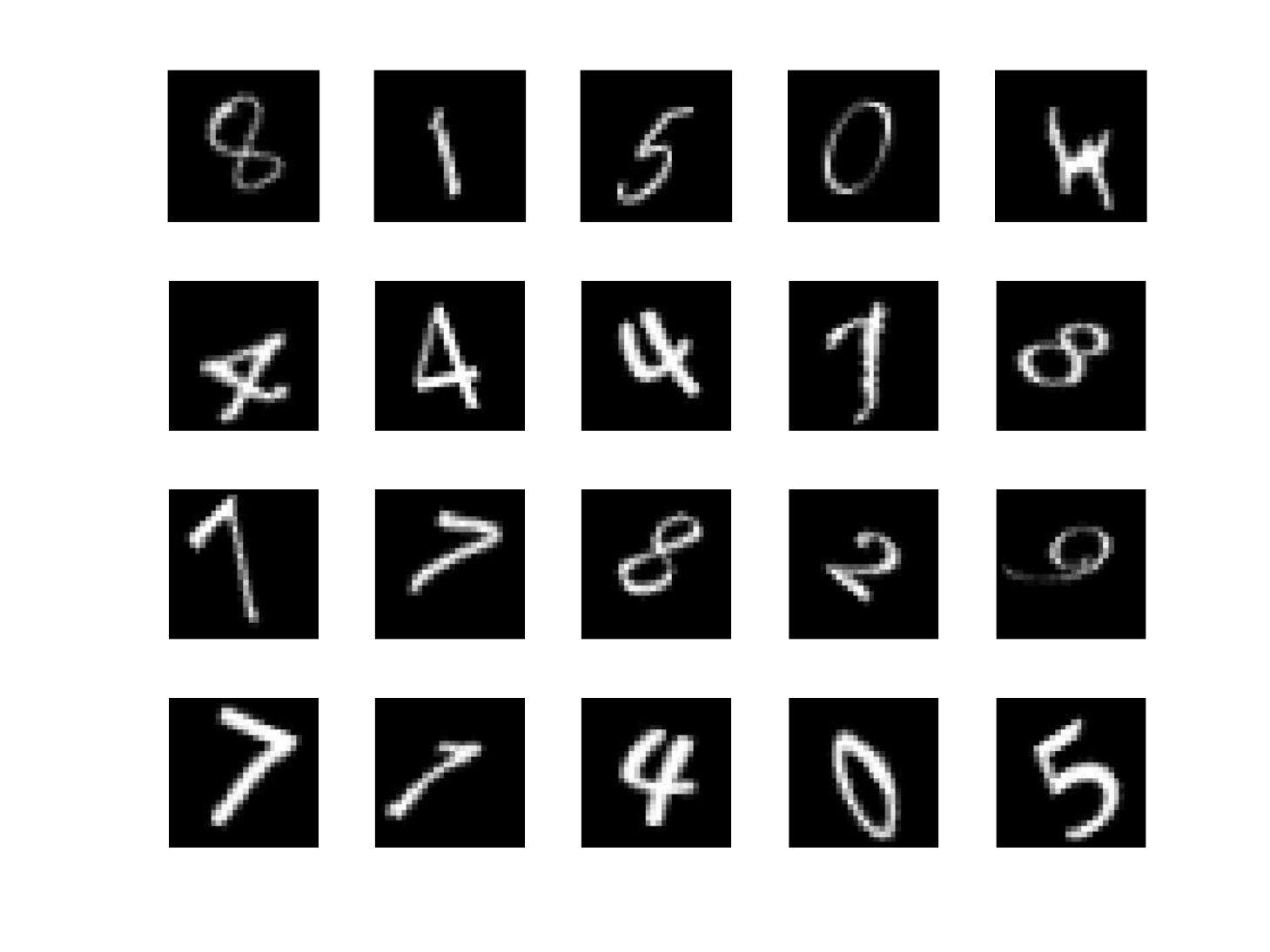}
		\caption{An example of 20 input images to the NN. \label{fig:inputNN}}		
	\end{center}		 		
\end{figure}

\begin{figure*}[t]
	\centering
		\resizebox{\textwidth}{!}{
		\begin{tikzpicture}
		\tikzstyle{connection}=[ultra thick,every node/.style={sloped,allow upside down},draw=\edgecolor,opacity=0.7]
		\tikzstyle{copyconnection}=[ultra thick,every node/.style={sloped,allow upside down},draw={rgb:blue,4;red,1;green,1;black,3},opacity=0.7]
		 	\draw (24,10,10) node [text=black, fill=\ConvColor, minimum width = 4cm]{Conv. Layer};
				\draw (24,9.5,10)  node [text=black, fill=\BatchNormColor, minimum width = 4cm]{Batch Norm. Layer};
				\draw (24,9,10)  node [text=white, fill=\SigmoidColor, minimum width = 4cm]{Sigmoid Layer};
				\draw (24,8.5,10)  node [text=white, fill=\PoolColor, minimum width = 4cm]{Avg. Pooling Layer};
				\draw (24,8,10)  node [text=white, fill=\DropoutColor, minimum width = 4cm]{Dropout Layer};
				\draw (24,7.5,10)  node [text=white, fill=\FcColor, minimum width = 4cm]{Fully Connected Layer}; 
		\node[canvas is zy plane at x=0] (input) at (-3,2,5) {\includegraphics[width=6cm,height=6cm]{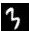}};
		
		\pic[shift={(0,0,0)}] at (0,0,0) 
		    {Box={
		        name=conv1,
		        caption= ,
		        fill=\ConvColor,
			opacity= 0.83,
		        height=28,
		        width=2,
		        depth=28
		        }
		    };
		
		\pic[shift={(0,0,0)}] at (conv1-east) 
		    {Box={
		        name=bn1,
		        caption= ,
		        fill=\BatchNormColor,
			opacity= 0.8,
		        height=28,
		        width=2,
		        depth=28
		        }
		    };
		
		\pic[shift={(0,0,0)}] at (bn1-east) 
		    {Box={
		        name=sig1,
		        caption= ,
		        fill=\SigmoidColor,
			opacity= 0.8,
		        height=28,
		        width=2,
		        depth=28
		        }
		    };
		
		\pic[shift={ (0,0,0) }] at (sig1-east) 
		    {Box={
		        name=pool1,
		        caption= ,
		        fill=\PoolColor,
		        opacity=0.8,
		        height=14,
		        width=1,
		        depth=14
		        }
		    };
		
		\pic[shift={(3,0,0)}] at (sig1-east) 
		    {Box={
		        name=conv2,
		        caption= ,
		        fill=\ConvColor,
			opacity= 0.83,
		        height=14,
		        width=2,
		        depth=14
		        }
		    };
		
		\draw [connection]  (pool1-east)    -- node {\midarrow} (conv2-west);
		
		\pic[shift={(0,0,0)}] at (conv2-east) 
		    {Box={
		        name=bn2,
		        caption= ,
		        fill=\BatchNormColor,
			opacity= 0.8,
		        height=14,
		        width=2,
		        depth=14
		        }
		    };
		
		\pic[shift={(0,0,0)}] at (bn2-east) 
		    {Box={
		        name=sig2,
		        caption= ,
		        fill=\SigmoidColor,
			opacity= 0.8,
		        height=14,
		        width=2,
		        depth=14
		        }
		    };
		
		\pic[shift={ (0,0,0) }] at (sig2-east) 
		    {Box={
		        name=pool2,
		        caption= ,
		        fill=\PoolColor,
		        opacity=0.8,
		        height=7,
		        width=1,
		        depth=7
		        }
		    };
		
		\pic[shift={(3,0,0)}] at (sig2-east) 
		    {Box={
		        name=conv3,
		        caption= ,
		        fill=\ConvColor,
			opacity= 0.83,
		        height=7,
		        width=2,
		        depth=7
		        }
		    };
		
		\draw [connection]  (pool2-east)    -- node {\midarrow} (conv3-west);
		
		\pic[shift={(0,0,0)}] at (conv3-east) 
		    {Box={
		        name=bn3,
		        caption= ,
		        fill=\BatchNormColor,
			opacity= 0.8,
		        height=7,
		        width=2,
		        depth=7
		        }
		    };
		
		\pic[shift={(0,0,0)}] at (bn3-east) 
		    {Box={
		        name=sig3,
		        caption= ,
		        fill=\SigmoidColor,
			opacity= 0.8,
		        height=7,
		        width=2,
		        depth=7
		        }
		    };
		
		\pic[shift={(3,0,0)}] at (sig3-east) 
		    {Box={
		        name=conv4,
		        caption= ,
		        fill=\ConvColor,
			opacity= 0.83,
		        height=7,
		        width=2,
		        depth=7
		        }
		    };
		
		\draw [connection]  (sig3-east)    -- node {\midarrow} (conv4-west);
		
		\pic[shift={(0,0,0)}] at (conv4-east) 
		    {Box={
		        name=bn4,
		        caption= ,
		        fill=\BatchNormColor,
			opacity= 0.8,
		        height=7,
		        width=2,
		        depth=7
		        }
		    };
		
		\pic[shift={(0,0,0)}] at (bn4-east) 
		    {Box={
		        name=sig4,
		        caption= ,
		        fill=\SigmoidColor,
			opacity= 0.8,
		        height=7,
		        width=2,
		        depth=7
		        }
		    };
		
		\pic[shift={(3,0,0)}] at (sig4-east) 
		    {Box={
		        name=do1,
		        caption= ,
		        fill=\DropoutColor,
			opacity= 0.8,
		        height=7,
		        width=2,
		        depth=7
		        }
		    };
		
		\draw [connection]  (sig4-east)    -- node {\midarrow} (do1-west);
		
		\pic[shift={(3,0,0)}] at (do1-east) 
		    {Box={
		        name=fc1,
		        caption= ,
		        fill=\FcColor,
		        opacity=0.8,
		        height=1.5,
		        width=1.5,
		        depth=1.5
		        }
		    };
		
		\draw [connection]  (do1-east)    -- node {\midarrow} (fc1-west);
		
		\end{tikzpicture}}		 	
	\caption{Architecture of the neural net used to estimate the angles of rotation. \label{fig:CNNArch}}		
\end{figure*}

As a third example, we use our approximate $1/x$ method is in a neural network (NN) for regression. For this task, we chose to use a NN to estimate the angles of rotation of the well known MNIST dataset~\cite{lecun-mnisthandwrittendigit-2010}. As a starting point of this example, we used the network described in~\cite{matlabCNN} and replaced all rectified linear unit (ReLU) layers with sigmoid activation functions, defined as
\begin{equation}
	f(x) = \frac{1}{1+e^{-x}}\,,
\end{equation}
so that the NN has division operations in its forward path. The network takes a $28\times 28$ black and white picture of the dataset as input (cf. Fig.~\ref{fig:inputNN}) and processes it through several convolution layers~\cite{Goodfellow-et-al-2016}. These layers extract features from the images useful for estimating its angle of rotation. Further, several batch normalization~\cite{DBLP:journals/corr/IoffeS15}, pooling~\cite{Goodfellow-et-al-2016}, and dropout~\cite{JMLR:v15:srivastava14a} layers are used to aid the NN in its task. As mentioned before, after each pooling, convolution, and batch normalization block, one activation function is used on the output of each of those blocks. In the original NN this was the ReLU function, which was replaced with a sigmoid activation to demonstrate the application of our reciprocal approximations in a neural network. This new architecture is shown in Fig.~\ref{fig:CNNArch} which was plotted using~\cite{haris_iqbal_2018}. The final output is obtained by a fully connected layer without activation functions to exploit the whole range of the output signal. In order to keep the implementation simple and to have a valid baseline, we trained the model on standard sigmoid layers using $1/x$ and then replaced the corresponding layers, after training, with approximations that make use of our $1/x$ approximation and, again for comparison, the $2^{-\lceil log_2(x) \rceil}$ approximation. To evaluate the different implementations with respect to their performance, we use the root mean squared error (RMSE) defined as
\begin{equation}
	\text{RMSE} = \sqrt{\frac{1}{N}\sum_{i=1}^{N}\left(y_i-\hat{y}_i\right)^2}\,,
\end{equation}


where the true value is given by $y_i$ and the NN's estimation is $\hat{y}_i$. We further define the accuracy as the percentage of predictions having an error of no more than $10$ degrees. The results are summarized in Tab.~\ref{tab:ResCNN}. It can be seen that our approximation yields in almost no change to the RMSE and accuracy. Further, the use of MPE leads to a slightly better performance. The approximation by $2^{-\lceil log_2(x) \rceil}$ results in an increase of approximately $175\%$ in terms of the RMSE and a decrease of more than 50\% in terms of accuracy.

\begin{table}
	\centering
	\caption{Performance of the ideal $1/x$ function and its approximations when used in the activation functions of a NN.}
	\begin{tabular}{l|l|l}
		            Algorithm             & RMSE     & Acc.\\ \hline
		              $1/x$               &  5.70    & 92.46\%\\
		    approx. $1/x$ w/ $C=26/9$     &  5.93    & 91.50\%\\
		approx. $1/x$ w/ $C=26/9$ and MPE &  5.76    & 92.16\%\\
		  $2^{-\lceil log_2(x) \rceil}$   &  15.72   & 42.10\%
	\end{tabular}
	\label{tab:ResCNN}
\end{table}

\section{Conclusion}
We presented an approximate variant of the reciprocal function, based on piecewise linear approximation, that can be efficiently implemented in digital hardware. We described a corresponding architecture that can be built using 
only combinatorial logic. Even when additional registers between the combinatorial sub-blocks are introduced, as it is common for practical applications, the presented approach can be implemented with $2$ to $4$ clock cycles maintaining high clock frequencies. We show synthesis results demonstrating the low area requirements and the high clock frequencies of the proposed design. We analytically described the error of the approach and show 
how to optimize a constant value used by the design. This way, different error behaviors of the design could be achieved and the constant was optimized accordingly. We finally present application examples that show practically negligible performance losses when using our proposed reciprocal function instead of the exact reciprocal function. 
\bibliographystyle{IEEEtran}
\bibliography{mybib}


\end{document}